\documentclass[aps,prx,reprint,longbibliography,superscriptaddress]{revtex4-2}
\usepackage{graphicx}
\usepackage{subcaption}
\usepackage{dcolumn}
\usepackage{bm}
\usepackage{float}
\usepackage{amsmath}
\usepackage{hyperref}

\begin{document}
	\title{Room Temperature Quantum Coherent Revival in an Ensemble of Artificial Atoms}

	\author{Igor Khanonkin}
	\email{ikhanonkin@technion.ac.il}
	\affiliation{Andrew and Erna Viterbi Department of Electrical Engineering and Russell Berrie Nanotechnology Institute, Technion, Haifa 32000, Israel}

	\author{Ori Eyal}
	\affiliation{Andrew and Erna Viterbi Department of Electrical Engineering and Russell Berrie Nanotechnology Institute, Technion, Haifa 32000, Israel}

	\author{Johann Peter Reithmaier}
	\affiliation{Institute of Nanostructure Technologies and Analytics, Technische Physik, CINSaT, University of Kassel, Kassel 34132, Germany}

	\author{Gadi Eisenstein}
	\affiliation{Andrew and Erna Viterbi Department of Electrical Engineering and Russell Berrie Nanotechnology Institute, Technion, Haifa 32000, Israel}

\begin{abstract}
	We report a demonstration of the hallmark concept of quantum optics: periodic collapse and revival of quantum coherence (QCR) in a room temperature ensemble of quantum dots (QD). Control over quantum states, inherent to QCR, together with the dynamical QD properties present an opportunity for practical room temperature building blocks of quantum information processing. The amplitude decay of QCR is dictated by the QD homogeneous linewidth, thus, enabling its extraction in a double-pulse Ramsey-type experiment. The more common photon echo technique was also invoked and yielded the same linewidth. Measured electrical bias and temperature dependencies of the transverse relaxation times enable to determine the two main decoherence mechanisms: carrier-carrier and carrier-phonon scatterings.
\end{abstract}

\maketitle

\section{Introduction}
	A two-level quantum system interacting with a resonant single mode of a quantized electromagnetic field is described by the famous Jaynes-Cummings model (JCM), which is the simplest quantum electrodynamic model that can be solved exactly. JCM was used as early as 1963 \cite{jaynes1963comparison} to demonstrate the quantum and semi-classical aspects of spontaneous emission. In 1980, Eberly et al. \cite{eberly1980periodic} solved the JCM in an exact manner for a weakly coupled interacting field which is initially in a coherent state and predicted a periodic recurrence of the quantum wave function, or QCR, following the decay of the upper state population (as per the classical Cummings collapse). Gea-Banacloche affirmed \cite {gea1990collapse} that during the collapse, the interacting atoms remain in a pure quantum state despite the fact that the atomic population inversion is zero. QCR was experimentally demonstrated first by Rempe et al. in 1987 in a one atom maser \cite{rempe1987observation} and then served as a measure of field quantization in cavity quantum electrodynamics starting from 1996 \cite{brune1996quantum}. 

	A quantized field is however not a necessary condition for QCR in a quantum two-level system but may be substituted for by a classical coherent field with a discrete energy spectrum. QCR was demonstrated with classical excitation fields in trapped atoms \cite{meekhof1996generation,andersen2003echo}, vibration states of gases \cite{bellini1997two, blanchet1998temporal}, excitons in GaAs quantum wells \cite{gobel1990quantum}, optical lattices \cite{raithel1998collapse}, and heavy – Fermi - liquid compounds \cite{wetli2018time}.  Furthermore, the collapsed wavefunction may recover only partially. This is termed fractional QCR which was introduced in 1989 by Averbukh et al. \cite{averbukh1989fractional} and demonstrated first in an atom-gas \cite{yeazellObservation1990,wals1994observation} and later in molecules \cite{vrakking1996observation,rudenko2006real, feuerstein2007complete}. 

	Thus far, revivals of quantum coherence in solids always required cryogenic temperatures, with no exception. In fact, room-temperature is unequivocally believed to diminish the effect altogether.

	Here we report on the demonstration of fractional QCR in a room temperature ensemble of semiconductor quantum dots (QDs) operating at the important wavelength range of 1550 nm. Our findings pave the way to practical elements for quantum information processing, communication and simulations where quantum states are controlled using compact semiconductor nanostructures operating at room temperature. The medium we use can utilize silicon photonics technology and is compatible with optical fiber telecommunication. The advantages of semiconductor nanostructures for quantum devices have long been recognized. Quantum devices and systems based on QDs \cite{waks2002quantum,li2003all,kroutvar2004optically,schwartz2016deterministic}, some used  in conjunction with photonic crystal waveguides and cavities \cite{faraon2008coherent,bose2012low,javadi2015single} were studied extensively as were special nano wires \cite{kitaev2001unpaired, rokhinson2012fractional, mourik2012signatures, das2012zero} which are also thought to be improved by combining them with QDs \cite{deng2016majorana}. Unlike our room temperature QCR demonstration, all those quantum elements operate exclusively at cryogenic temperatures.
	
	\begin{figure}[b]
		\centering
		\includegraphics[width=8cm]{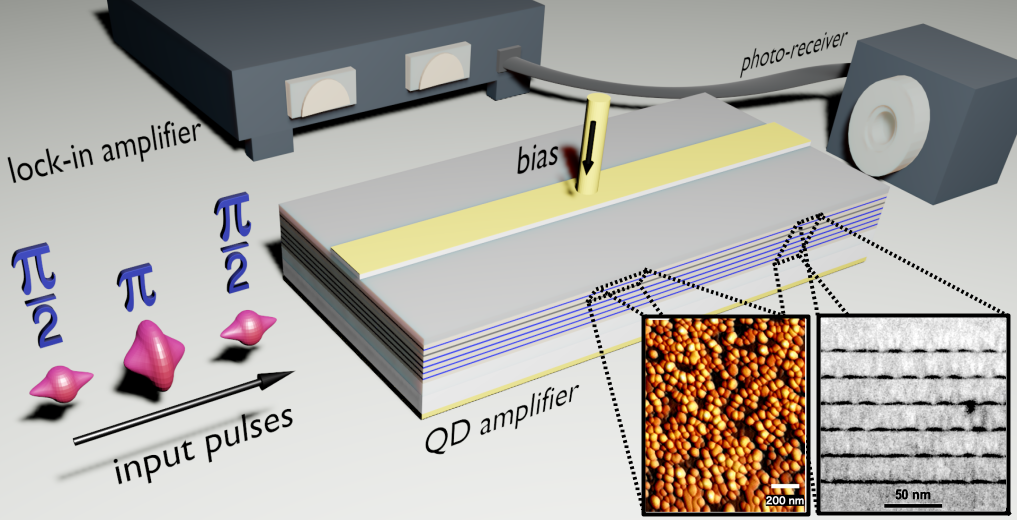}	
		\caption{Schematic of the experiment. $90~fs$ wide pulses were split in two and three for the $\pi/2$ area Ramsey pulse pair and the $\pi$ area rephasing pulse, respectively. The intensity of the second Ramsey pulse was measured at the output of the  InAs/InP QD optical amplifier using a lock-in scheme. The inset shows an atomic force microscope image of a single layer of self-assembled QDs as well as a high resolution transmission electron microscope cross section of the six QD layer structure.}
		\label{Fig_setup} 		
	\end{figure}
	
	The platform we used for the QCR experiment is a $1.5~mm$ long InAs/InP QD optical amplifier operating at room temperature and driven electrically to the gain regime in the 1550 nm wavelength range. We employed the Ramsey configuration with two time delayed $90~fs$ pulses. The excitation pulse induces coherent interactions with a number of discrete groups of QDs (modes) which are a subset of the inhomogeneously distributed QDs. These modes constitute an effective inhomogeneous linewidth that determines an effective inhomogeneous transverse relaxation time $T_{2}^{*}$. This set of modes plays the role of the quantized photon numbers as in \cite{eberly1980periodic} and of elements in a discrete classical spectrum \cite{averbukh1989fractional}, namely, the modes interfere constructively to induce QCR. The decay of the consecutive revival amplitudes is governed by the homogeneous transverse relaxation time, $T_2$ whose extracted value was confirmed by a Ramsey-echo experiment with a three pulse measurement. Finally, we present measured dependencies of the extracted $T_2$ and $T_{2}^{*}$ values on the carrier density and the temperature from which we determine the decoherence mechanisms.
	
\section{Experimental conditions} 
	
	The experimental configurations we employed is schematically depicted in Fig. \ref{Fig_setup}. Pulses with a duration of $90~fs$, measured by FROG, with a repetition rate of $40~MHz$ were split in two or three for the Ramsey and Ramsey-echo experiments, respectively. The delay between pulses can be controlled with a sub femto-second resolution, achieved using accurate DC motorized linear stages. The Ramsey pair and the rephasing pulses were cross-polarized to TE and TM, respectively. The second Ramsey pulse was chopped, and its average power was measured using a photo-receiver and a lock-in amplifier at the QD waveguide output. To ensure complete pulse separation at the input, the minimum nominal temporal delay between co-polarized pulses was $600~fs$. 
	
	The active region of the gain medium we used comprised six layers of high-density, $6 \cdot 10^{10}~cm^{-2}$ InAs QDs grown by molecular beam epitaxy in the Stranski - Krastanow mode.  The amplifier was fabricated as a $2~\mu m$ wide, $1.5~mm$ long ridge waveguide whose end facets were anti-reflection coated. The inset in Fig. \ref{Fig_setup} depicts an atomic force microscope image of a single QD layer as well as a high resolution transmission electron microscope cross section of the six layer structure. The QDs exhibit record uniformity characterized by the photo-luminescence linewidth at $10~K$ which was $17~meV$ and $26~meV$ a single QD layer and for the six layer stack, respectively \cite{banyoudeh2015high}. The emission power of the measured electro-luminescence spectra increases with applied bias but the spectral shape remains unchanged \cite{khanonkin2017ultra}. This ensures that for all bias levels, the excitation pulse interacts with the same QDs and the pulse area changes linearly with bias.
	
	During the delay between the Ramsey pair of pulses, the quantum state of the system evolves in a periodic manner (in the rotating frame approximation) with the classical frequency of the atom transition on which an exponential decay, representing damping, is superimposed. The second Ramsey pulses probes directly those damped Ramsey interference fringes and consequently yields the inhomogeneous transverse relaxation time. The Ramsey-echo technique employs an additional excitation pulse (called the rephasing pulse) which is launched in the middle of the Ramsey pair. The rephasing pulse reverses the phases accumulated across the inhomogeneously broadened ensemble so that the system coherence is determined solely by the homogeneous linewidth \cite{allen1987optical}.
	
	The Ramsey method requires that the excitation pulses have areas of $\pi / 2$. The QD optical amplifier is a unique interaction medium in that the pulse area cannot be determined unequivocally. Upon propagation along the $1.5~mm$ long active waveguide, the pulse experiences gain as well as nonlinear spectral modifications \cite{karni2015nonlinear}. This change of pulse properties means that only an effective pulse area can be defined. The  $\pi / 2$ area of the TE polarized Ramsey pulses was set by optimizing their input energy to obtain the largest possible Ramsey interference contrast at a given bias. This ensures that the effective pulse area is as close as possible to $\pi / 2$. In the present experiment, the first and second input pulse energies that yield the closest area to $\pi/2$, when the amplifier is driven at $7.15~kA/cm^2$, were $7.5~pJ$ and $5.6~pJ$, respectively. The intensity of the $\pi$ area rephasing pulse should be $\sqrt{2}$ larger than that of the $\pi / 2$ Ramsey pulses. However, due to the polarization dependent dipole moment (determined from gain (G) measurements), the intensity is larger. It was set to be $\sqrt{2} \cdot G_{TE}/G_{TM}$ larger than that of the first Ramsey pulse. 

	\begin{figure*}[htb]
		\centering
		\includegraphics[width=17.8cm]{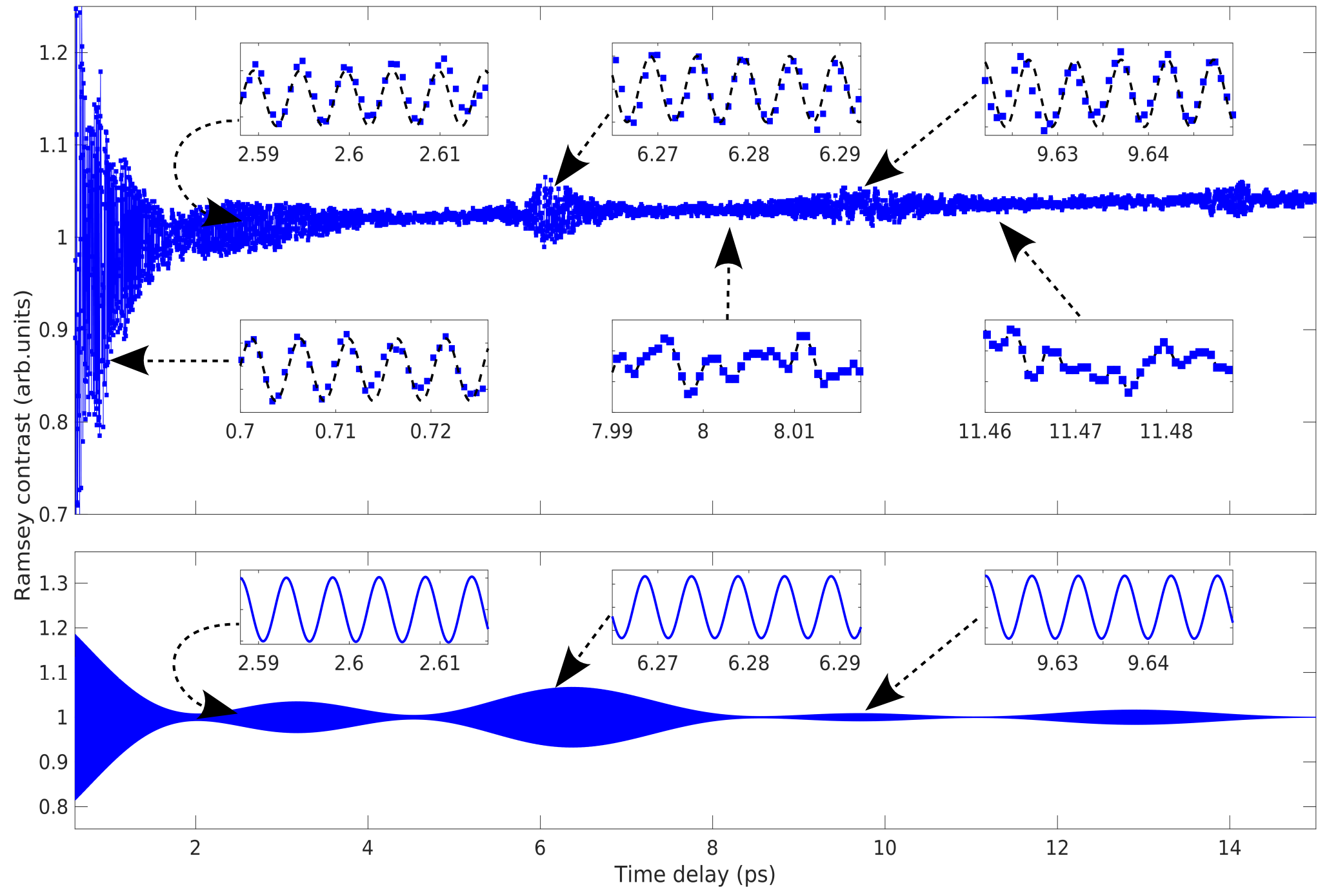}
		\caption{Top: Measured normalized Ramsey contrast at a current density of $7.15~kA/cm^2$ revealing fractional QCR. The insets show for clarity the Ramsey fringes with a periodicity of $\approx 5.1~fs$ corresponding to the QDs transition energy at $1535~nm$ and noise in between the revivals. The initial decay is dictated by the effective inhomogeneous transverse relaxation time, $T_{2}^{*}$, while the fractional QCR decays according to the homogeneous transverse relaxation time, $T_2$. Bottom: Normalized analytical solution of five excited uncoupled modes each with a dephasing time of $T_2=4.64~ps$, spectrally located close to $1535~nm$. The calculated Ramsey fringes are also shown in the inset for clarity.}
		\label{Revival_15ps}
	\end{figure*}

\section{Experimental and theoretical results} 

	The measured intensity of the second Ramsey pulse at the amplifier output for a current density of $7.15~kA/cm^2$ is presented in the upper part of Fig.\ref{Revival_15ps} for a time span ranging from $600~fs$ to $15~ps$.  The initial Ramsey contrast decay which is governed by $T_{2}^{*}= 1.22~ps$ resembles the Cummings collapse. At later times, four periodic cycles of coherence, representing fractional QCR that originate from the interference of a few excited discrete modes, each with a slightly different transition frequency, are clearly observed. The insets show imprints of Ramsey interference fringes with periodicity of $\approx 5.1~fs$, corresponding to one optical cycle at the QDs gain peak, $1535~nm$. In between the revivals, the response is just noise. 

\begin{figure}[htb]				
		\includegraphics[width=8.6cm]{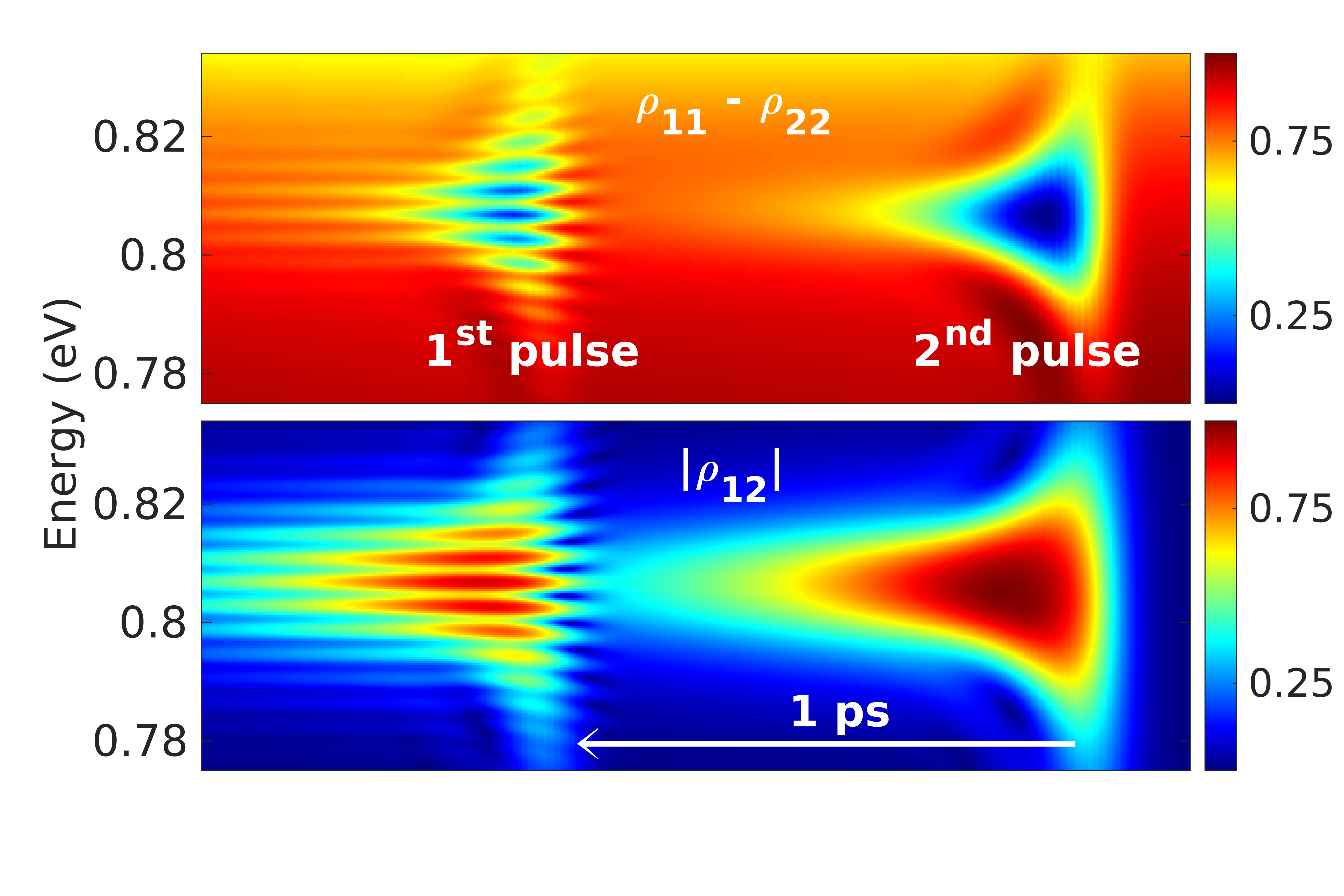}
		\caption{Simulated Ramsey - type experiment. The spectral dependence of the normalized QDs population inversion (upper trace) and coherences (lower trace) upon Ramsey pulse pair propagation with a $1~ps$ mutual delay for a few hundred $\mu m$ along the amplifier.}
		\label{Simulation} 
\end{figure}

	The spectral placement of the pulses relative to the QD gain peak as well as pulse amplitude and spectral irregularities determine the set of modes to be excited in a coherent manner that can be sensed in a two pulse Ramsey - type experiment. The particular modes participating in the interaction, for the experimental conditions leading to  Fig.\ref{Revival_15ps}, were determined using a comprehensive model of a propagating pulse in a QD amplifier  \cite{capua2013finite}. The model solves Maxwell and Schr$ \ddot{o}$dinger equations in the presence all nonlinear and nonresonant interactions \cite{karni2015nonlinear,khanonkinRamsey2018}. The simulation enables to map the population inversion ($\rho_{11}-\rho_{22}$) and the coherences ($\rho_{12}$) anywhere along the amplifier and at all wavelengths with $\rho_{ij}$ being elements of the density matrix. Fig. \ref{Simulation}  describes normalized population inversion and coherences of the QD ensemble when two Ramsey pulses, separated from each other by $1~ps$, have propagated a few hundred $\mu m$ along the amplifier. The oscillatory nature shows clearly five dominant modes. The spectral signature of the five modes changes somewhat as the pulses propagate along the amplifier but the general shape is maintained. 

	The measured revival pattern can be reconstructed with those five interacting modes as shown in the lower part of Fig. \ref{Revival_15ps}. An analytical time resolved response is used: $\exp(-t/T_2) \cdot \sum_{k}  a_k  \sin (2 \pi t / t_{k})$, where the five interacting modes are considered to be uncoupled with weighted amplitudes $a_k=\{1/3, 1/2, 1, 1/2, 1/3\}$ and oscillation periods of $t_k=(5.109+k \cdot 0.004)~fs$. Each mode is broadened to $1.79~meV$, leading to a dephasing time $T_2=4.64~ps$. The analytical calculation fits the experimental result very well.

\begin{figure}[htb]				
		\includegraphics[width=8.6cm]{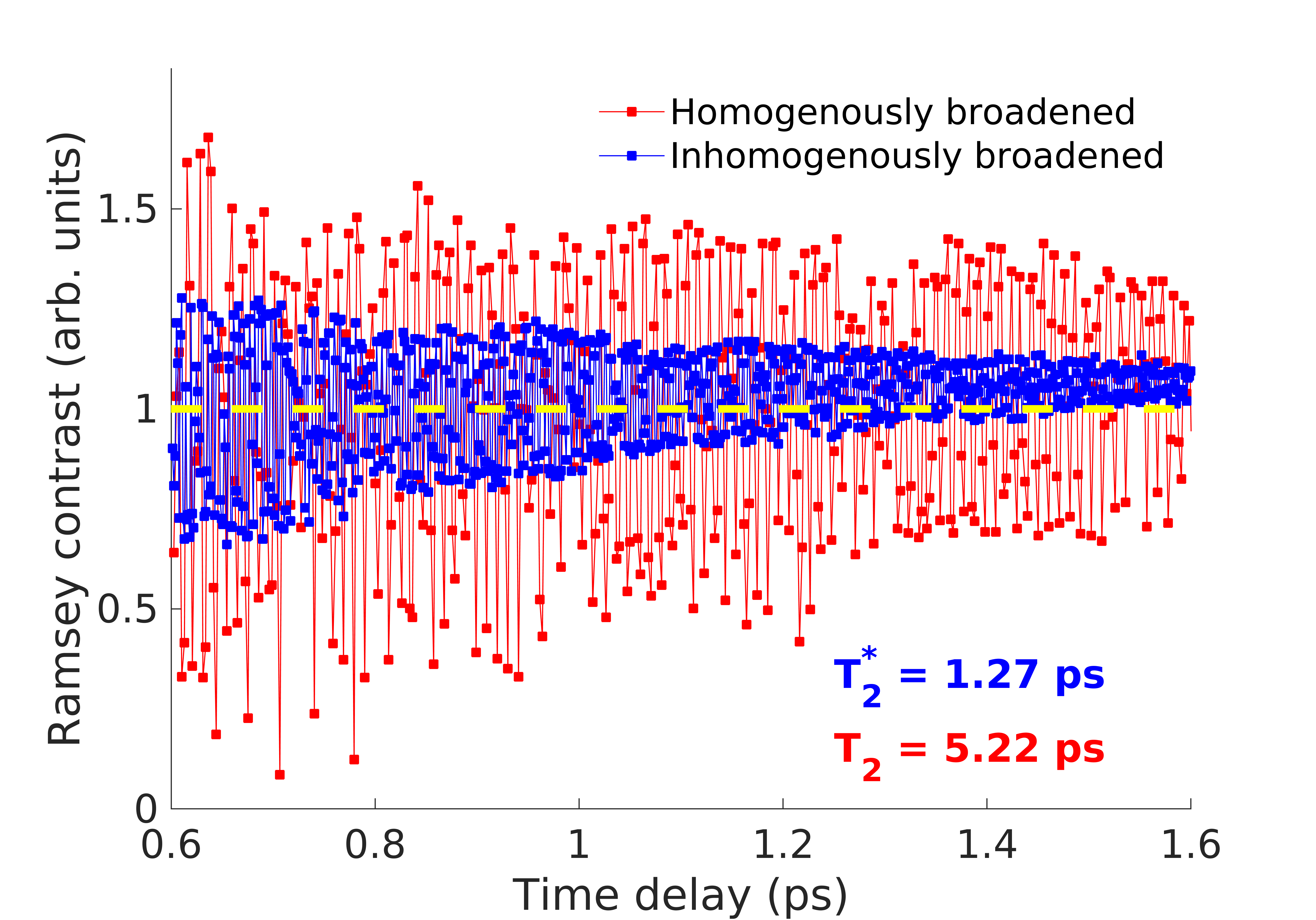}
		\caption{Contrast of Ramsey interference fringes measured at a current density of $4.7~kA/cm^2$ in two pulse  (blue trace) and three pulse (red trace) experiments. The coherence decays according to $T_{2}^{*}=1.27~ps$ extracted from the blue trace. The three pulse Ramsey - echo experiments yields $T_2=5.22~ps$. The yellow line of the contrast equal 1 is shown for eye guiding.}
		\label{Figure_4} 
	\end{figure}

	The initial exponential decay in Fig.\ref{Revival_15ps} provides the effective inhomogeneous transverse relaxation time $T_{2}^{*}$, while the amplitude reduction of the periodic fractional QCR occurs on a time scale of the QD dephasing time, $T_2$. To confirm the latter, we performed a Ramsey-photon echo measurement. Fig.\ref{Figure_4} compares the Ramsey contrast, measured in the standard Ramsey technique (blue trace) with that measured in the induced photon echo configuration (red trace) for a current density of $4.7~kA/cm^2$. Clearly, the photon echo technique prolongs the QDs coherence substantially. From the two type of experiments we fit the measured Ramsey contrasts according to $exp(-2t/T_{2}^{*})$ and $exp(-4t/T_{2})$ \cite{yajima1979spatial} and extract the two time constants $T_{2}^{*} =1.27~ps$ and $T_2=5.22~ps$, respectively. For a larger bias, the dephasing time shortens and at $7.15~kA/cm^2$ it reduces to $4.64~ps$. 
	
	The experimental traces are not symmetric relative to the time axis.  This asymmetry stems from nonlinear absorption of the first Ramsey pulse. Charge carriers absorbed during the incoherent nonlinear interaction relax to the QD ground states on a pico-second time scale and provide additional gain \cite{khanonkin2017ultra}. This problem was overcome by averaging the upper and lower exponential decays of the fringe envelopes. As the Ramsey coherent interference has a time scale of a few femto-seconds, the contrast is unaffected by the incoherent nonlinear propagation.

\begin{figure}[htb]				
		\includegraphics[width=8.6cm]{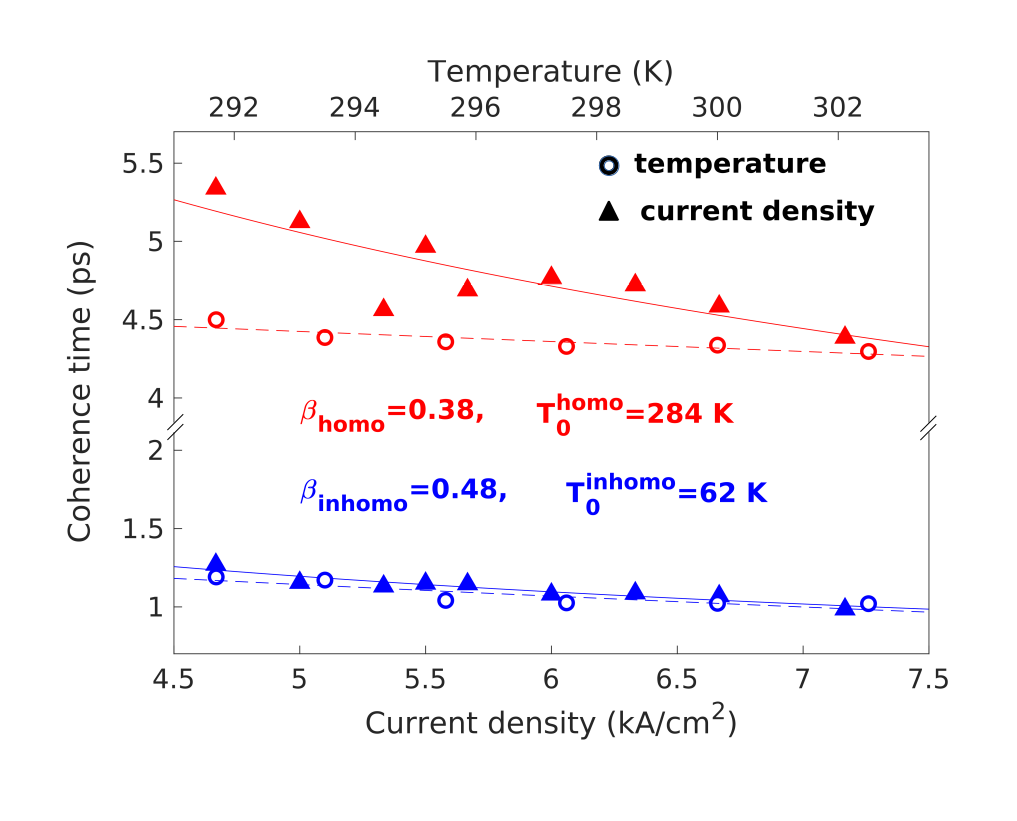}
		\caption{$T_2$ (red trace) and $T_{2}^{*}$ (blue trace) dependencies on current density and temperature representing carrier - carrier and carrier - phonon scatterings. The current density dependencies follow a power law $ \propto n^{-\beta}$ yielding for the homogeneous broadened $\beta_{homo}=0.38$ and inhomogeneous broadened $\beta_{inhomo}=0.48$. The temperature dependencies are fitted with an exponential function $ \propto exp(-T/T_0)$ yielding $T_0^{homo}=284~K$ and $T_0^{inhomo}=62~K$.}
		\label{Figure_5} 
	\end{figure}

	Operation at room temperature affects the system decoherence significantly and therefore it is crucial to understand, in detail, the mechanisms causing decoherence. There are two main mechanisms of decoherence \cite{shah2013ultrafast}: carrier - carrier scattering originating from Coulomb interaction between QD confined charge carriers and the surrounding area, and carrier-phonon scattering involving acoustic and longitudinal optical phonons. These two mechanisms were examined by measuring $T_{2}^{*}$ and $T_2$ in double and triple pulse experiments for various applied carrier densities (n) and temperatures (T). Fig. \ref{Figure_5} summarizes the extracted characteristic times with a predicted scaling of $ \propto n^{-\beta}$ and $ \propto exp(-T/T_0)$ for the bias and temperature dependencies, respectively. $T_{2}^{*}$ was found to be less dependent on the carrier density as compared with $T_2$. The extracted fitting coefficients are $\beta_{homo}=0.38$ and $\beta_{inhomo}=0.48$. For carrier-carrier scattering, these coefficients are known for InAs in the bulk ($\beta^{bulk}_{homo}=0.3$) and in quantum wells ($\beta^{well}_{homo}=0.5$) \cite{hugel2000dephasing}. An exact comparison with the obtained value for the QDs is difficult due to the different dimensionality and due to the fact that in the QD structure, some of the carrier-carrier scatterings take place in the quantum well - like high energy barrier. Nevertheless, the obtained value of $\beta_{homo}$ is in the right range and sheds light on the origin of the scattering events. The value of $\beta_{inhomo}$ cannot be compared to that in a quantum well or in the bulk since those are homogeneous systems.

	Fig.\ref{Figure_5} describes also an almost constant $T_2$ in the relatively narrow, measured temperature range while $T_{2}^{*}$ appears to be strongly dependent on temperature; the extracted fitting coefficients are $T_0^{homo}=284~K$ and $T_0^{inhomo}=62~K$, respectively. Since at room temperature the separation between the first excited energy state (p-shell) and the ground state (s-shell) in the InAs QDs is $\approx 60~meV$ \cite{khanonkin2017ultra} which equals roughly the energy of two longitudinal optical (LO) phonons ($\approx 30~meV$ \cite{buchner1974raman}), relaxation by carrier-phonon interaction is probable as observed in the case of GaAs QDs with multi LO-phonon resonances whose homogeneous broadening was shown to be temperature dependent above $200~K$ \cite{bayer2002temperature}. 
	
	To conclude we have demonstrated the important quantum optics phenomenon of fractional QCR in a room temperature semiconductor QD based optical amplifier driven by an electrical bias to the gain regime in the wavelength range of 1550 nm. QCR manifests itself as periodic recurrences of the quantum coherence which originate from constructive interference between excited homogeneous QDs subgroups. The coherent interference of the QDs modes persists up to the characteristic homogeneous time $T_2$, thus, enables its extraction without the need to invoke the photon echo technique. 
	
	In the present experiment, the properties of the excitation pulse spectrum determine the number of interacting modes to be five, and this leads to the particular revival sequence shown in Fig. \ref{Revival_15ps}. Furthermore, the revival pattern may be modified in a controlled manner using a pulse shaping technique. 

	Semiconductor QDs are the most attractive solid state platform for quantum information processing, communication and simulation. QD amplifiers have actually been shown to maintain their properties at temperatures as high as $100~^{0} C$ \cite{eyal2017static}. We envision a host of future quantum devices, possibly integrated with silicon photonics, leading to compact solutions in quantum systems. The prospect of operating at elevated temperatures advances their practical implementations.
	
\begin{acknowledgments}

	The authors are grateful to Prof. Moti Segev, Prof. Ariel Kaplan, and Prof. Eric Akkerman for important suggestions during the course of this research. I. K. thanks the Helen Diller Quantum Center and the Jacobs Foundation for financial support. This research was partially funded by the Israel Science Foundation, grant number 1504/16.
	
\end{acknowledgments}
	
\bibliography{bibliography}

\end{document}